\documentclass[seceq]{ptptex}
\usepackage{wrapft}

\pubinfo{Vol.~11?, No.~?, 200?}

\title{
Virtual Laboratories
}

\author{
Piet \textsc{Hut}$^{1,}$\footnote{E-mail: piet@ias.edu}
}

\inst{
$^1$Institute for Advanced Study, 1 Einstein Drive, Princeton, NJ 08540, USA
}

\recdate{
Tanabata, 7/7/2006}

\abst{
At the frontier of most areas in science, computer simulations play
a central role.  The traditional division of natural science into
experimental and theoretical investigations is now completely outdated.
Instead, theory, simulation, and experimentation form three equally
essential aspects, each with its own unique flavor and challenges.
Yet, education in computational science is still lagging far behind,
and the number of text books in this area is minuscule compared to the
many text books on theoretical and experimental science.  As a result,
many researchers still carry out simulations in a haphazard way,
without properly setting up the computational equivalent of a well
equipped laboratory.  The art of creating such a virtual laboratory,
while providing proper extensibility and documentation, is still in
its infancy.  A new approach is described here, Open Knowledge, as an
extension of the notion of Open Source software.  Besides open source
code, manuals, and primers, an open knowledge project provides
simulated dialogues between code developers, thus sharing not only the
code, but also the motivations behind the code.
}

\begin{document}

\maketitle

\section{Introduction}

Computer simulations began more than fifty years ago as straightforward
extensions of theoretical pen-and-paper calculations.  Even in those
days, computers were faster than human calculators by several orders
of magnitude.  This already created totally new opportunities, such as
the possibility for astrophysicists to do virtual laboratory experiments.

Whereas astronomy is the oldest of the physical sciences, it was the
one branch of science that could not put its subject matter, stars,
in a lab.  Unlike other environmental sciences, like meteorology and
geology, they could not even touch their subject matter, and the best
they could do was to observe the ongoing experiments in the universe
at enormous distances, without any form of control.  But when Martin
Schwarzschild and his collaborators used John von Neumann's computer
in Princeton in the early fifties, suddenly it became possible to put
a whole star in a virtual laboratory, and to watch it evolve over time
scales of millions and billions of years.

In the half century since then, computers have increased in speed by
more than ten orders of magnitude.  As a result, all areas of science
have been affected.  And no wonder: a quantitative increase in speed
over human pen-and-paper abilities of fifteen orders of magnitude is
reflected in an enormous qualitative advance as well.  Just open any
science journal and look at any article therein, and the chances are
high that such an article could not have been written using pure
thought.

However, when you walk into a bookstore, library, or office of a
working scientist, the overwhelming majority of books still reflects
the way science was done in the middle of the previous century.
Sure, they mention the results of computer simulations, but by and
large they do not give precise instructions for obtaining those results.
And if they do, the instructions apply to some very specific small part
of the virtual laboratory involved, never to the whole challenge of
setting up and running a complete computational environment.

The same strange time warp is visible in undergraduate education.
Students learn how to solve differential equations, they master vector
and tensor algebra, and they learn how to apply all those tools to
theoretical calculations in quantum mechanics or other areas.  If they
are lucky, they may learn some algorithms for numerically solving
differential equations, and they may be exposed to one or more
computer languages.

In addition to all that, one might expect that students would be
taught how to perform reasonably large-scale computational
experiments in such a way that they would become familiar with the
nuts and bolts of setting up, running, and analyzing such virtual
laboratory experiments -- tasks that go far beyond a familiarity with
numerical algorithms and the grammar of a few computer languages.
Sadly, that is not the case.

If you are a student ready to start a real piece of research, you are
faced with a basic choice: find black box software that you can run
without really understanding how it works; or find an older student
who can help you to get acquainted with the details of existing code
to the extent to the extent that you can modify it and/or write new
code for your own purpose.  The way in which insight is communicated
is basically medieval.  In order to help build a cathedral, you just
start working with an existing crew, and over time oral knowledge
sinks in, while you get hands-on experience in the actual work.

This mode of operation puts students at a tremendous disadvantage.
For many of them, there just may not be any individual nearby who can
provide the oral knowledge needed for a given project.  And even for
those students with access to people with experience, they and their
informants have to spend very many hours sitting and working together
before the necessary information is finally shared.  This type of
guild-based information sharing may have its charming aspects, but it
is no longer appropriate in a democratic approach to science in an age
of globalization.

Sharing knowledge with peers is an essential aspect of science.
And working collaborations between senior and junior scientists
provide great opportunities for younger people to learn the ropes
from their seniors in actual research settings.  However, it is a
waste of time and energy to let more senior people show the very
basic steps toward working in virtual laboratories over and over
again.  As a result, most students after a while avoid asking too
many questions.  They then have no choice but to reinvent many
wheels.  Most wheels produced this way may be just adequate for the
immediate job at hand, but do not lend themselves to being extended
to wider usage.

The worst effect of all this is that students are effectively trained
to cobble things together, without ever acquiring the skills to
develop a vision for building or even just maintaining and extending
a whole computational lab.  No wonder that many researchers are still
using coding styles and tools that are decades out of date.

\section{Open Source}

In some fields of science, the number of researchers is large enough
and the funding is lavish enough that commercial companies find it
profitable to develop packages that are specifically tailored to the
type of simulations that are most often carried out.  In most areas,
however, the size of the user base would be too small to make
commercial development of packages an option.  In those cases,
scientists have no other choice but to build their own simulation
environments.

\subsection{Code}

Even in those case where packages are available, a typical package
will not allow you to do precisely what you had in mind.  And the
longer you are experimenting, the more likely it is that you will
want to modify the package, to allow it to perform somewhat different
experiments, or somewhat different ways of analyzing the results of
the experiments.  In such cases, it is very helpful if the canned
software is of the `open source' type, where users have full access
to the source codes.

But this is only the first step.  Whether a scientist gets the full
source code for an existing simulation package from a friend and
colleague, or from a commercial package, in both cases it will be
an uphill battle to figure out how to extend such a code to include
new features.

\subsection{Comments}

For starters, most codes are written in a hurry.  Commercial codes are
always late, and commercial code writers feel their managers breathing
down their neck.  Under such pressure, the chances are slim that the
resulting code will be commented very carefully.  For codes developed
by scientists themselves, the situation is often worse.  For one thing,
academic researchers have their own time pressures in the form of dead
lines for finishing a PhD, getting their next postdoc, getting tenure,
applying for a new grant, getting their students to finish their PhD
in time; it never ends.  For another, scientists typically receive no
training at all in properly commenting their codes.

But let us assume that a scientist is lucky and gets a code from a
colleague that is commented in such a clean way that it is possible
to figure out what each subroutine does and what the meaning is of
the variables used.  That would be a second step, and it would be
very helpful indeed.  Still, if we are dealing with a large code
with several tens of thousands of lines and hundreds of subroutines,
even utter clarity of individual subroutines may not be sufficient
to let us understand the structure of the code as a whole.

\subsection{Manual}

Let us presume that our lucky scientist gets even luckier, and gets
presented with some form of manual.  Commercial products are likely to
come with a manual, and even if the source code is available (which
is often not the case) such a manual will be a great help when trying to
understand the workings of a large code.  In contrast, codes written
by scientists rarely come with a detailed manual.  But if a manual can
be found, this will comprise a third step toward insight in the code,
after openness of the source and good code comments.

In practice, though, almost any manual in existence reflects the state
of the code at some point in the past, and cannot be relied upon to
tell you what the latest version of the code is really doing.  Or more
likely, some parts of the manual are only a little bit out of date,
while other parts that have not been revised recently, may be far more
out of date.  As a result, a typical manual is not even internally
consistent, completely apart from the problem of not reflecting the
code as it exists right now.

But let us move on in our wishful thinking.  Imagine, if you can, a
large but very clean code, well written, well commented, and well
documented with a wonderful manual.  You are presented with such a
code, and you want to start doing research with it.  Now what?  How
do you get started?  What are the right parameters to use?  How can
you start to do a small simulation, which will return its results
quickly, just to test the waters?  And how will you know whether the
numbers that come out of such a test simulation are correct?

\subsection{Primer}

A good manual describes the functionality of the code, including the
layout of the different parts, the relations between them, and the way
each part can be invoked.  However, the type of questions raised above
are not part of what a manual is supposed to cover.  Instead, we would
like to have a primer, a type of introduction to how to use the code,
complete with practical examples, ways of testing the results, pointers
to the literature, discussions of the scientific relevance, and so on.

A good primer is the fourth step toward being able to explore a code,
and get acquainted with it.  But even with a good primer, it may be
very hard to extend an existing code.  When presented with a legacy
code with more than ten thousand lines, how are you going to be sure
that you don't break something when you modify or add some pieces?
And even if you can avoid breaking something, wouldn't it be nice to
have a sense of how to make a new piece fit in a functional way with
the old stuff?  If you could only know what the original writers knew
. . .

\subsection{Conversations}

In other words, wouldn't it be nice to have a number of leisurely
conversations with the writers of the code?  If you had those people
at hand, and if they were willing to take the time to tell you what
was on their mind when they wrote the code, wouldn't that be great?

If they would tell you which approaches they tried out at first,
before settling on what was enshrined in the legacy code, you could
avoid making the same mistakes that they did.  Or you might notice
that one or more ideas which they labeled as a mistake actually do
offer interesting possibilities when you extend them a bit.  In either
case, it could save you a large amount of time if you could talk
with them, and in addition it would give you many new ideas that are
not evident from code, comments, manual, or primer -- even in the almost
ridiculously optimistic assumption that all those four would be available.

But . . . there is yet a different obstacle.  Imagine that you get
free access to the writers of the code, and you can spend as much time
as you like picking their brains.  What you would find, very quickly,
is that they themselves would have forgotten the answers to most of
your questions.  Because they almost certainly made no notes, or at
best very sketchy notes, it will be impossible to reconstruct what
actually happened during the writing of the code.

\subsection{Time Machine}

What you really want is not to talk with the people writing the code
as they are now.  Instead, you'd like to take a time machine and travel
back to meet their younger versions, while they were busy writing their
code.  If you could only be a fly on the wall, watching them at work,
or better even, if you would be able to read their mind while they were
working at their computers, wouldn't that be something!

Of course, like with any fairy tale or parable, if you really got what
you were wishing for, it would be impractical.  If someone would have
made a movie of the process of code writing, complete with mind reading
ability, it might take you a year to watch all the installments of the
movie.  Writing code is a creative process, and it will be very hard
to anticipate which part of a year-long tape you should watch to get
an answer to a particular question.  Creative ideas don't come in
systematic frameworks, and code writers are likely to jump around while
making modifications, in ways that cannot be anticipated.

\section{Open Knowledge}

So what's the solution?  If hiring the code writers as personal
consultants wouldn't work, and if time machines and mind readers
would not be enough, how can we possibly hope to transfer knowledge
from working scientists writing code to other scientists wanting to
extend the code years or decades later?

\subsection{Parallel Universe}

Let's keep fantasizing just a bit more.  As long as we are allowed
to travel in time and read minds, why not assume that we can travel
to a parallel universe, where scientists are a whole lot smarter
than the ones we normally encounter here on Earth.  Imagine that
you would meet two or three scientists in that other universe who
were busy writing just the sort of legacy code you were trying to
decipher.

Since they are a lot smarter, they will not waste as much time on
debugging as we tend to do, and they will come up with the right
idea far more quickly than anyone we know.  Still, they are not
infinitely smart, and they, too, will have to grope in the dark,
as part of the creative process of software writing.  Imagine that
they are smart enough to develop the code in a way that produces
a movie that is much shorter than a year, but that they are still
limited enough that they will make the sort of mistakes that we
would make.  The only difference is that they are just more nimble
and quick in getting past all their initial errors.

Finally, let us assume that these ideal code writers have yet one
more perfect feature.  At regular intervals, they will take a break,
and they will talk to each other about what they have done so far,
what they will do next, and how this fits into the grand scheme of
things.  So in addition to being amazingly clever and amazingly fast,
they are also amazingly well organized.  Yes, they make mistakes,
and yes, they have to try this, that and the other thing, before
figuring out how to take the next step in their code development,
but after each groping in the dark they take some time off to map
carefully what they have found so far.

Early on in our story, we made increasingly unlikely assumptions about
the availability of clean code, clear comments, detailed manuals, and
wonderfully instructive primers.  After that, we lost touch with reality
by introducing time machines, mind reading equipment, and access to
exceedingly gifted researchers in parallel universes.  However, in
doing so, we did come up with a solution for the problem we posed at
the outset: how to convey knowledge from the scientists who originally
wrote a large code to the scientists who later want to modify that code.
The main idea is to be a fly on the wall in a parallel universe, while
watching unrealistically clever code writers write the code you're
trying to figure out.

\subsection{Simulating Writing Simulation Codes}

Apart from all those unrealistic details, the solution itself is fine.
And it may be the only solution!  Any other solution that I have ever
come across falls way short in comparison to this solution.

So what are we to do, without access to parallel universes?
Well, the whole story has been about computer simulations.
Why not simulate access to a parallel universe, populated with
benevolent clever scientists ready to help us?!

Here is the basic idea.  If we write a computer code that is aimed at
performing simulations, we can perform a simulation of the process of
writing the simulation code, in real time.  In this way, we are time
sharing with ourselves.  While thinking about writing a piece of code,
we take some time off, and write a page or so about our thinking, in
a condensed way.  After that, while writing bits and pieces of code,
we again take some time off at regular intervals to write one or
more pages, each time, about what it is that we are coding. 

\subsection{Three Central Ideas}

There are three central ideas in simulating ourselves while writing a
simulation code.  First, we have to flag the central points in each
day of work, hits and misses,  successes and failures, new ideas that
lead to something and new ideas that don't.  Keeping a note pad at
hand, or more likely an open computer file, while working on the code
is a good discipline for making sure to catch whatever is important.

Second, we have to condense the whole learning process by one or two
orders of magnitude, in order to avoid producing an endlessly boring
story.  Each important point should feature in the simulation of the
thought process of the code writer, but most of the fluff should be
taken out.

Third, and most important: the simulation of the code writing should
not be allowed to lag behind the actual code writing.  It may be very
tempting to press on with the real code, and to leave the simulation
to be written the next day or the next week.  Especially in complex
situations where your fingers are itching to try this and that to see
what works, you may be hard pressed to take time off to simulate
yourself.  But it is precisely in those situations that the simulation
is most needed, for later readers and for yourself as well, when
you've forgotten how and why you've written what when.  The main point
is: once you begin to postpone the simulation of the writing process,
you are likely to fall behind more and more, until you just can't
reconstruct that process any more.

\subsection{Dialogues}

As for the format to choose for simulating the thought processes going
into code writing, the most natural style is a dialogue.  To start
with, it is much more interesting to read a dialogue than a long
string of thoughts of a single person.  In addition, a dialogue format
offers the possibility to introduce different characters.  For example,
one person may tend to go for quick and dirty results, while the other
one may prefer a more formal approach.  Left to their own devices,
either approach tends to become too extreme, and a good simulated
dialogue gives the possibility to trade off these two tendencies,
showing how a well written code can be produced from a balance between
striving for clarity of logic and a pragmatic wish for rapid results.

Apart from the practicality of using a dialogue for the simulated code
writing, there are many advantages to actually doing the real code
writing with two people.  Recently, various books have appeared
describing the advantages of so-called pair programming, also known in
some form as extreme programming.  In my experience, the fact that a
day of work will take up two person-days this way is more than offset
by the increase in quality of the code, and in many cases in the speed
of code writing as well.  It is far more likely that two people choose
a fruitful direction early on, and similarly, a mistake by one person
will often be caught right away by the other.  Last, but not least,
writing code with someone else can be a lot of fun, as a learning
experience as well as having a good time.

The result of all these considerations is a variation on an Open
Source approach to software writing that I have developed over the
last few years in collaboration with Jun Makino.  We call our approach
Open Knowledge\cite{rf:1}, to indicate that not only our codes are
freely available, but also the knowledge that went into the code
writing process and that was developed while the code was written.

\section{Earlier Projects}

\subsection{Modest Beginnings}

Before describing our Open Knowledge project, it may be useful to
describe the context within which Jun Makino and I decided upon our
new approach to simulate the generation of simulation code.  The main
motivation for doing so was our dissatisfaction with the traditional
approaches that we had been engaged in, together with lessons learned
from various alternatives that we explored during the last two decades.

Our earliest coding experiences were with relatively small codes.
In the early eighties I studied the gravitational three-body problem
in great detail, and the codes I wrote for that purpose were typically
only a few thousand lines long, including the preparations of initial
conditions and the analysis of the results of a three-body scattering
experiment.  In the mid eighties Jun Makino wrote a one-dimensional
climate modeling code, which also was less than a few thousand lines.

As a result, both of us got a real taste of the joy of being able to
design a whole system from scratch, while having complete control
over all the details, with the possibility of extending any part where
needed, as soon as it is needed.  This stands in sharp contrast with
the experience of many students whose first research experience is
with a legacy code that they cannot possibly hope to understand in all
detail.

In other words, when we got started, Jun and I failed to learn the
standard lesson that you're not supposed to understand a scientific
computer program in detail.  This has colored the rest of our career
in computational science, for better or worse.

\subsection{Aarseth Codes}

Both of us moved to the general gravitational $N$-body problem during
the mid to late eighties, and in doing so we encountered two
surprises: a happy surprise tied to the particular field of stellar
dynamics, and an unhappy surprise related to scientific software in
general.

The happy surprise was that we found that one individual, Sverre
Aarseth, had set the tone for a friendly and helpful collaborative
atmosphere in stellar dynamics.  Not only did he make his codes freely
available but in addition he offered his help generously to anyone
using his codes.  Frequently, he would even extend his code for the
special purpose of helping someone to explore a new parameter regime
for which the code had not yet been tested.  In this way, Sverre's codes
became more and more robust, in a type of evolutionary adaptation to a
wide variety of research programs.

The not-so-happy surprise was that we realized that the clarity of
coding, that we had become used to in small applications, did not
carry over to large programs.  In all large scientific codes that
we came across, not only in astrophysics but in all fields of
computational science, we felt like immigrants from the country
side, visiting a big city for the first time.  In our small towns,
we knew all the streets and landmarks by name and by sight, but in
the big cities we got hopelessly lost.  While some big cities
had a rudimentary map, many cities had not even that.  And no city
we explored had an in-depth guide book or collection of detailed
historical descriptions, to explain how the city had grown into its
present shape.

In fact, Sverre Aarseth's code documentation was significantly better
than what was available for most codes.  In the mid eighties, he wrote
a very helpful summary\cite{rf:2}, which made it possible for Jun and
me to get a quick overview of the main algorithms used.  But even so,
it was a momentous task to go through the whole code and try to
understand all its details, something that very few people have ever
gotten around to do.

As an aside, concerning modern developments: recently Sverre published
a more detailed book\cite{rf:3} in which he describes the main
ingredients that have gone into his code building in the last half
century.  While this book makes it far easier to understand the
overall structure of the code and the basic ideas behind it, a
traditional book size limit of a few hundred pages makes it impossible
to provide more than a sketch for each of the many key ideas that have
been gathered over the decades.  In various places, a particular
choice made in Sverre's codes is described by a sentence such as
``after considerable experimentation.''  For a typical user this may
be fine, but for someone wanting to try alternative options, there is
no choice but to repeat this considerable amount of work from scratch,
to gain similar experience.

Sverre's book is mostly directed at collisional stellar dynamics,
where the interactions between individual particles, representing
stars, are important.  In contrast, many areas of stellar dynamics
fall in the category of collisionless stellar dynamics, where the
particles only sample phase space.  These particles play two roles:
they act like smoke in a six-dimensional wind tunnel, introduced to
visualize the flow, and at the same time they sample the gravitational
field in order to compute the dynamics of that flow.  The most popular
and well documented code of this type is Gadget\cite{rf:4}, written by
Volker Springel.

\subsection{Nemo}

Returning to the mid eighties, in 1985 Josh Barnes and I developed our
tree code\cite{rf:5} for performing large-scale $N$-body calculations in
$N\log N$ time, rather than $N^2$.  At first we used our code directly
to perform many exploratory calculations as well as production runs for
particular scientific applications.  However, we soon felt the need to
get more organized, rather than having to deal with zillions of files
distributed haphazardly over many different directories.  We then
stumbled upon the idea of developing a type of virtual laboratory, to
help us set up, carry out, and analyze the results of our simulations.

We gave our software environment the name Nemo, because our programs
resided in a shared directory, owned by nobody in particular.  It
seemed natural to called the package `nobody', but since that name was
already reserved on our system, we chose the Latin translation
instead.  Some of our ideas were developed in a series of discussions
with Gerald Sussman and with Peter Teuben, who soon took over the main
role of `captain Nemo'\cite{rf:6}, while Josh developed a new version,
called Zeno\cite{rf:7}.  Some of our ideas were published in the
proceedings of a conference I organized in 1986 in Princeton, The Use
of Supercomputers in Stellar Dynamics\cite{rf:8}, in a popular
version in Scientific American\cite{rf:9}, and a few years later in
a conference on Celestial Mechanics\cite{rf:10}.

Nemo consists of an environment with a toolbox, in the form of a large
set of moderate-size programs which perform individual tasks,
involving manipulations of $N$-body systems.  Examples are the
generation of initial conditions, the selection of subsystems, the
addition, translation and rotation of systems, and the representation
of systems in the form of various two-dimensional projections of the
full phase space information.  The system was developed under UNIX,
with an architecture based on pipes and filters, to allow the user to
chain together a number of operations in a single command line.  With
this type of flexibility, it became very easy to engage in small
interactive exploratory calculations before setting up large runs in
batch mode.

\subsection{Starlab}

Nemo quickly developed into an efficient environment for galactic
dynamics, where the emphasis was on collective effects, rather than on
the dynamics of individual particles.  In fact, a single particle in a
simulation of the collisions of two galaxies has no direct physical
meaning: it samples the phase space distribution of stars in a Monte
Carlo version, since the total number of stars involved, on the order of
$10^{12}$, is far too large to model on a one-to-one basis.  Therefore,
great care is taken in such calculations to suppress as much as
possible the non-physical local interactions between the particles.

In contrast, simulations of the stellar dynamics of dense stellar
systems\cite{rf:11} require the explicit modeling of individual stars.
In a globular cluster, for example, single stars may transform into
double stars or triples, or they may be born as such, and these
multiple systems play an important role in the energy budget of the
star cluster as a whole.  Much of the algorithmic challenges of star
cluster dynamics stems from the difficulty modeling the physical
effects of close encounters.  The numerical challenges are formidable:
spatial length scales vary over more than fifteen orders of magnitude,
from the diameter of a neutron star to the size of the cluster, and
time scales vary over more than twenty-one orders of magnitude, from
the sound crossing time in a neutron star to the age of an old star
cluster.

In order to simulate star cluster dynamics, a new package was developed,
built on the previous experience we had gained with Nemo.  Starting in
1989, while on a sabbatical at Tokyo University, I wrote the first
version of Starlab, which was based on a more flexible type of data
structure than that used in Nemo.  Like Nemo, Starlab is a collection
of loosely coupled programs, linked at the level of the UNIX operating
system, that share a common data structure, and can be combined in
arbitrarily complex ways to study the dynamics of star clusters and
galactic nuclei.  A couple of years later, Steve McMillan joined the
development effort of Starlab, and in 1992, Jun Makino also joined us,
at which point we decided to switch from C, as used in Nemo and the
earlier Starlab, to C++\cite{rf:12}.  At that time, Simon Portegies
Zwart began to develop his SeBa stellar evolution code\cite{rf:13},
as an integral part of Starlab.

In Starlab, the basic data structure for grouping the stars together
is a tree (rather than an array, as is the case in Nemo).  This allows
a natural way to group bound stars in binaries and higher-order multiples,
and to allow proper numerical treatment for unbound stars during unusually
close encounters.  In addition, the use of Unix style pipes is encouraged
more than it was in the case of Nemo, because in Starlab data are not lost
when not recognized by a particular module; unlike in Nemo, unrecognized
data are simply passed on to the next module.  Finally, the data structures
have been designed from the start to allow multi-physics applications, such
as the additions of stellar evolution and stellar hydrodynamics.

\section{The Art of Computational Science}

\subsection{Moving Stars Around}

Twelve years after the first Starlab programs were written, in 2001,
Jun Makino and I looked back at the many design decisions that were
made developing the package, especially during the first few years.
Our main motivation was to experiment with alternative choices of data
structures, and to develop alternative algorithms for the treatment of
the interaction between the local and global dynamics of stars in a
star cluster or galactic nucleus.  We were especially interested in
the treatment of neighboring particles close to a multiple star
system, in the so-called perturber list.  The treatment of perturbers
in Sverre Aarseth's NBODY6 code, as well as in the Kira code, the main
engine of the Starlab environment, is far from transparent.
Therefore, we wondered whether it would be possible to come up with
more simple and clear versions, at not too large a performance
penalty.

We were considering starting to write yet another package, following
Nemo and Starlab, but at that time we had enough experience to predict
what was likely to happen.  Such a package would probably start out in
a clear and clean way, only to become more and more tangled internally
under later pressure to perform reasonably efficiently, for a variety
of research projects.  Therefore, before embarking on another major
enterprise, we asked ourselves the general question of how we could
avoid the fate of other packages that we were familiar with, and had
contributed to ourselves.

One early tentative conclusion we reached was: research = education,
at least in computational science.  The best way to write a readable
program, complete with comments, manual, primer, and additional
background material, is to treat it as an educational project.
The benefit of such an approach is not only that it will become much
easier to train students to join such a project.  At least as
important is the fact that the researchers themselves, a few years
later, will be able to let themselves be educated by the younger
versions of themselves who wrote the educational material.  Practice
has taught us how essential it is to make detailed and organized notes
of the considerations that go into the writing of a code: if you don't
do this quickly, you are bound to have to repeat the `considerable
amount of experimentation' that went into the original code writing in
the first place.

We therefore decided to start off by writing an introductory text for
simulations in stellar dynamics, Moving Stars Around\cite{rf:14}.
Using the C++ language, we started with a hands-on description of
simple integration algorithms, such as forward Euler and leapfrog,
applied to the two-body problem, followed by the use of a Hermite
scheme for the general N-body problem, with applications to the cold
collapse of a group of particles, and subsequent binary star
formation.  We have used this 250-page manuscript as introductory
material in two summer schools\cite{rf:15}, where we received many
positive reactions of the students.  There clearly seems to be a large
gap in the market for introductory material for computational science,
and our manuscript was one of the few publications that addressed the
problem of giving hands-on guidance.

\subsection{Kali and Maya}

That first volume, Moving Stars Around, which we published as open
source on the web, was written in the form of a dialogue between three
students.  The use of a dialogue form seemed to be the most natural
vehicle for giving the reader a sense of how software is developed in
practice.  Both of us had already experimented with the medium of
dialogues\cite{rf:9,rf:16}, and we had gotten enough positive feedback
to be encouraged to continue those experiments.

In the fall of 2003, after completing our first volume, we embarked on
a far longer project, aimed at developing the Kali code, a
state-of-the-art set of computer programs to simulate dense stellar
systems, together with a software environment in which to use such a
code.  We chose the name from the Sanskrit word {\it kali}, meaning
{\it dark}, as in the {\it kali yuga}, the dark ages we are currently
in according to Hindu mythology.  The same word also occurs in the
name {\it Kali}, for the Hindu Goddess who is depicted as black.  The
term {\it dark} seemed appropriate for our project of focusing on
forms of tacit knowledge that have not been brought to light, so far,
and perhaps cannot be presented in a bright, logical series of
statements.  Instead, we expect our dialogues to carry the many less
formal and less bright shades of meaning that pervade any craft.

For the software environment in which to run the Kali code, we chose
the name Maya, which seemed fitting for two reasons, one connected
with Middle America and one with India.  The Maya culture was very
good at accurate calculations in astronomy.  And the word {\it maya}
in Sanskrit has the following meaning, according to the Encyclopedia
Britannica: "Maya originally denoted the power of wizardry with which
a god can make human beings believe in what turns out to be an
illusion."  Indeed, a simulation of the heavens is something virtual,
an illusion of sorts, and a considerable feat of wizardry.

In order to allow optimal flexibility in rapid prototyping, we decided
to switch from C++ to a scripting language, and our choice, for the
time being at least, fell on Ruby.  By the summer of 2004, we had
written more than a thousand pages of dialogues, spread over a dozen
volumes, which we put up on the web in open source form as soon as
they were written, together with the computer programs described in
them.  During the subsequent two years, it has been very encouraging
to us to hear from many readers with very diverse backgrounds that
they have found that material already useful, even though it is only a
small start for what will become a project that we expect to span tens
of thousands of pages.

\subsection{A Four-Dimensional Look at the $N$-Body Problem}

One of the novel ideas in the Kali code is to view the $N$-Body
Problem as more than just a time-dependent configuration of points
in three dimensional space.  Instead, from the start we take a
four-dimensional spacetime view, in which the $N$-Body Problem
becomes an $N$-Orbit-Segment Problem.  An essential difference
between the two pictures is that the latter allows for asynchronism,
where the events in space and time where forces are calculated no
longer need to be correlated between different world lines.

For many applications in stellar dynamics, all particles can share
the same time step size, the value of which can either be chosen
beforehand, or determined adaptively during a run.  However, for dense
stellar systems, with their inherently huge ranges of length and time
scales, we are forced to abandon the option of letting all particles
use the same time step.  Instead, we have no choice but to use
individual time step values, where each particle has its own value for
the time step size, different from most other particles\cite{rf:3}.

Individual time step schemes are not normally treated in text books on
numerical solutions of differential equations, and its use implies a
number of tricky issues.  Whenever a particle wants to update its
position and velocity, it has to determine the net acceleration that
all other particles exert on it, even though the positions and
velocities of most other particles are not known at the time desired
for the update.  In order to know where all other particles are at that
time, the original particle has to poll the others, to ask them where
they would expect be at that time, and each of the other particle will
then use a form of extrapolation to provide that information.

When this process is described in the usual three-dimensional way of
looking at a stellar dynamics system, it all seems quite messy and
potentially confusing.  Therefore, Jun Makino and I have started to
experiment with ways of writing an individual time step code from a
four-dimensional point of view, in spacetime rather than space as our
arena.  The basic ingredients are no longer particles but world lines,
each one belonging to one particle moving in time.  At any given time,
some part of each world line has been computed, and that orbit segment
can be used to obtain all the necessary information to let both that
segment and other orbit segments grow further in time.

To our pleasant surprise, this switch to a four-dimensional
perspective has made it far easier to write more flexible code,
including the introduction of individual algorithms, besides the use
of individual time step values.  We have developed early versions of
the Kali code where each particle has a choice of
algorithm:\cite{rf:17} some particles can be integrated with a
Runge-Kutta code, others with a Hermite code, yet others with the
original Aarseth multi-step algorithm, and so on.  It is a testimony
to the flexibility of a scripting language like Ruby, combined with
the natural view offered by a world line perspective, that we were
able to write individual-algorithm codes rather quickly and easily.

As an unexpected bonus of our four-dimensional orientation, it turned
out to be relatively easy to write a time-symmetric block time step
version of an $N$-body code\cite{rf:18}, something that never had been
done before.  Here block time steps are choices for individual time
steps in a binary fashion: each particle has a time step that is equal
to the largest permissible time step, or to a time step that is
smaller by an integer power of two.  The key idea in our
paper\cite{rf:18} is to apply an iterative time symmetrization
procedure to a whole timespan or era, a block of spacetime that
includes all of space and an amount of time equal to an integer number
of longest permissible time step values.

\subsection{The Near Future}

Currently, we have taken a break from our Kali code development, in
order to rewrite our original `Moving Stars Around' introduction.  In
the new version, we have added far more introductory material, to make
that volume even more accessible for students with relatively little
background, as well as for students who want to refresh their
understanding of the basic concepts of numerical orbit integration.

The new volume includes Ruby, rather than C++, as our main teaching
device, and it will also include far more sophisticated plotting and
visualization tools, something that was largely lacking in the
original 2003 version.  We have made the first 100 pages of the new
book available on our web site\cite{rf:19}, and we will add more
material soon.  As always, feedback will be greatly appreciated.

When we return to develop our Kali code further, time symmetric block
time steps will be a high priority for us, as well as the development
of regularization techniques for close encounters.  Simultaneously, we
plan to develop our Maya environment further, to allow sophisticated
semi-automated experimentation in a virtual laboratory setting, along
the lines of what we envisioned already more than twenty years
ago\cite{rf:8, rf:9}.

\section{Conclusions}

The history of the use of simulations in science is only half a
century old, a factor of ten shorter than the use of theory and
experimentation in modern science.  Therefore, it is perhaps no
surprise that we are still in an initial phase where we are groping
for the right approach toward building and testing virtual tools and
designing virtual laboratories\cite{rf:20}.

During the same period, in the second half of last century,
experimentation has moved from the activity of individuals and small
groups to larger and larger collaborations, sometimes containing hundreds
or even thousands of experimental scientists.  Currently, the enormous
increase in computer speed is precipitating a similar transition in
scientific simulations.  Where individual simulators could still write
and maintain their own codes, not too long ago, this is becoming almost
impossible to do nowadays.  The emphasis is rapidly shifting toward
group efforts, and as a result, the sociology of computational science
is rapidly shifting as well.

When the task of developing and maintaining a complex code or set of
codes is distributed over several researchers, the problems of
documentation and communication becomes even more acute\cite{rf:11},
especially when the researchers are not located in the same place, or
even the same country.  There is an urgent need to face up to these
problems, and it has become very clear that the traditional approaches
to writing some comments in a code and adding a few summary pages as
an appendix to scientific articles is no longer sufficient.

Most likely, half a century from now we will have found a more mature
and stable way to distribute the various responsibilities in virtual
laboratories over groups of computational scientists.  But the only
way to arrive at such solutions is to encourage a wide range of innovative
approaches, to try to see which ones are most practical and efficient.

The open knowledge approach\cite{rf:1}, developed by Jun Makino and
me, and described in the current paper is one such approach, which
follows our earlier efforts in the form of Nemo\cite{rf:6} and
Starlab\cite{rf:12} (for a long list of links to other codes,
see the Nemo web site\cite{rf:21}).  In short, open knowledge provides
added value to the basic idea of an open source approach to sharing
code, by adding a simulation of the process of developing codes for
scientific simulations.  It is our hope that others will develop some
completely different approaches, and that we can engage in friendly
competitions in order to understand and further improve the merits of
each approach.

\section*{Acknowledgments}
I thank Hans-Peter Bischof, Stan Blank, Doug Hamilton, Jun Makino,
Ernie Mamikonyan, Cole Miller, Bill Paxton and Peter Teuben for their
comments on the manuscript.
I thank Masao Ninomiya for inviting me to present this talk at the workshop.
I thank Shin Mineshige for inviting me to visit the Yukawa Institute for
Theoretical Physics, at Kyoto University, where I wrote the current paper.
This work was supported in part by the Grants-in-Aid of the Ministry
of Education, Science, Culture, and Sport  (14079205, 16340057 S.M.), by the
Grant-in-Aid for
the 21st Century COE ``Center for Diversity and Universality in Physics''
from the Ministry
of Education, Culture, Sports, Science and Technology (MEXT) of Japan.

\end{document}